\documentclass[aps,twocolumn,floats,prd,nofootinbib,showpacs,showkeys]{revtex4} 
\usepackage[dvips]{graphicx} %
\usepackage{epsfig,amsmath}
\usepackage{amssymb}
\usepackage{rotate}
\usepackage{color}
\usepackage{bm}

\DeclareFontFamily{OT1}{pzc}{}
\DeclareFontShape{OT1}{pzc}{m}{it}%
             {<-> s * [1.10] pzcmi7t}{}
\DeclareMathAlphabet{\mathscr}{OT1}{pzc}%
                                 {m}{it}

\newcommand{\be}{\begin{equation}}
\newcommand{\ee}{\end{equation}}
\newcommand{\bea}{\begin{eqnarray}}
\newcommand{\eea}{\end{eqnarray}}
\def\ba#1\ea{\begin{align}#1\end{align}}

\newcommand{\refeq}[1]{Eq.~(\ref{eq:#1})}          
\newcommand{\refeqs}[2]{Eqs.~(\ref{eq:#1})--(\ref{eq:#2})}          
          
\newcommand{\reffig}[1]{Fig.~\ref{fig:#1}}

\newcommand{\vs}{\nonumber\\}       
\newcommand{\refsec}[1]{\S~\ref{sec:#1}}          
          
\newcommand{\p}{\mathscr{p}}

\renewcommand{\v}[1]{\mathbf{#1}}

\newcommand{\vn}{\bm{\nabla}}
\newcommand{\vx}{\v{x}}
\newcommand{\vz}{\v{z}}
\newcommand{\vk}{\v{k}}

\newcommand{\vy}{\v{y}}

\renewcommand{\vr}{\v{r}}

\newcommand{\Wt}{\widetilde{W}}

\newcommand{\<}{\langle}
\renewcommand{\>}{\rangle}

\newcommand{\eps}{\varepsilon}
\renewcommand{\d}{\delta}

\newcommand{\D}{\Delta}

\newcommand{\rhob}{\overline{\rho}}

\newcommand{\iMpch}{\,h~{\rm Mpc}^{-1}}

\newcommand{\Msunh}{\,M_{\odot}/h}

\newcommand{\Om}{\Omega_m}

\newcommand{\s}{\sigma}

\newcommand{\M}{\mathcal{M}}
\newcommand{\F}{\mathcal{F}}
\newcommand{\fNL}{f_{\rm NL}}
\newcommand{\perm}{\mbox{perm.}}
\newcommand{\dng}{\hat{\d}}

\begin{document}

\title{Halo Clustering with Non-Local Non-Gaussianity}

\author{Fabian Schmidt and Marc Kamionkowski}
\affiliation{California Institute of
     Technology, Mail Code 350-17, Pasadena, California  91125}

\begin{abstract}
We show how the peak-background split can be generalized to
predict the effect of \emph{non-local} primordial
non-Gaussianity on the clustering of halos.  Our approach is
applicable to arbitrary primordial bispectra.  We show
that the scale-dependence of halo clustering predicted in the
peak-background split (PBS) agrees with that of the
local-biasing model on large scales.  On smaller scales,
$k\gtrsim 0.01\iMpch$, the predictions diverge, a consequence of
the assumption of separation of scales in the peak-background
split.  Even on large scales, PBS and local biasing do not generally
agree on the amplitude of the effect outside of the high-peak limit.  
The scale dependence of the biasing---the
effect that provides strong constraints to the local-model
bispectrum---is far weaker for the equilateral and
self-ordering-scalar-field models of non-Gaussianity.  The bias
scale dependence for the orthogonal and folded models is weaker
than in the local model ($\sim k^{-1}$), but likely still strong enough to be
constraining.  We show that departures from scale-invariance of
the primordial power spectrum may lead to order-unity
corrections, relative to predictions made assuming
scale-invariance---to the non-Gaussian bias in some of these
non-local models for non-Gaussianity.  An Appendix shows that a
non-local model can produce the local-model bispectrum, a
mathematical curiosity we uncovered in the course of this
investigation.
\end{abstract}


\date{\today}

\maketitle

\section{Introduction}
\label{sec:intro}

An active effort to seek departures from Gaussianity in primordial
cosmological perturbations is currently under way
\cite{Bernardeau:2001qr}.  The motivation stems from the desire
to learn more about inflation \cite{inflation,NGReview}.  The
simplest single-field slow-roll (SFSR) inflation models predict
that any departures from Gaussianity should be undetectably
small \cite{localmodel}.  However, theorists generally consider these
to be toy models, or effective theories, and there should
be, at some point, phenomena that depart from the predictions of
these simplest models.  Many physics models for inflation
\cite{larger,curvaton,Dvali:1998pa}, or
alternatives/additions to inflation
\cite{topdefects,Figueroa:2010zx}, do in fact make predictions
for non-Gaussianity at a much higher level than those of SFSR inflation.

It has been proposed to search for primordial non-Gaussianity
through a variety of techniqes, including cluster, void, and
high-redshift-galaxy abundances and properties
\cite{abundances,Verde:2000vq}.
Primordial non-Gaussianity had long been sought with
large-scale-structure surveys \cite{LSS}, but the sensitivity of
these surveys had generally been weak compared with that due to
the CMB \cite{Verde:1999ij,limits,Komatsu:2010fb}.
It has recently been shown, however, that the effects of primordial
non-Gaussianity may be amplified in the clustering of
dark-matter halos
\cite{Dalal:2007cu,Slosar:2008hx,Matarrese:2008nc,DesjacquesSeljak}.  The number
density of halos is
determined by the threshold overdensity for gravitational
collapse.  In
a region with a long-wavelength overdensity, the upward
fluctuation required for collapse is reduced and the
abundance of objects thus higher.  However, in non-Gaussian
models, the local power spectrum (or density-fluctuation
amplitude) may also be affected by the local long-wavelength
overdensity.  If so, this provides an additional dependence of
the local abundance of halos on the long-wavelength
density field.  In the local-model of non-Gaussianity, this
leads to a rapid increase in the biasing of halos on large
scales.  Null searches for this large-scale enhancement
place constraints to primordial non-Gaussianity that are
competitive with those from the CMB \cite{Slosar:2008hx,Xia:2010yu}.

Still, the local model is just one of many possible models for
non-Gaussianity, and it is natural to inquire how other forms of
non-Gaussianity, described by other bispectra, may affect
biasing.  This question was tackled in Ref.~\cite{Verde:2009hy}
using local Lagrangian biasing (LLB) \cite{Matarrese:1986et} to
model the clustering of halos.  
However, different analytic descriptions of biasing that make similar 
predictions at lowest order may differ at higher order \cite{Catelan:2000vn}, 
and so it is important to investigate different biasing models to
understand better the model predictions and theoretical
uncertainties.  In this paper, we calculate the clustering of halos for a
general primordial bispectrum using the peak-background split
(PBS).  Our approach generalizes that in
Refs.~\cite{Dalal:2007cu,Slosar:2008hx} (see also \cite{GiannantonioPorciani})
and complements the local-bias calculation of Ref.~\cite{Verde:2009hy}. 

As examples, we apply our results to the equilateral \cite{Equil},
folded \cite{folded}, orthogonal \cite{Senatore:2009gt}, and
self-ordering-scalar-field (SOSF) \cite{Figueroa:2010zx}
bispectra, as well as the local-model bispectrum
\cite{Luo:1993xx,Verde:1999ij}.  
We find that the scale-dependence of the biasing---the
effect that leads to strong constraints to the local-model
bispectrum where the bias correction goes as $k^{-2}$ ---disappears for the equilateral and
self-ordering-scalar-field models.  While the scale dependence
for the orthogonal and folded models is weaker than in the local
model ($\sim k^{-1}$), it may still be strong enough to be constraining.  We
also find that small departures from scale-invariance in the
primordial power spectrum may lead to large corrections to the
non-Gaussian bias for some of these non-Gaussian models,
especially for the equilateral and SOSF models.

We show that while the PBS prediction for the scale-dependence of halo 
clustering for general bispectra agrees
with the LLB predictions at large scales, they differ at small
scales, a breakdown of the separation of scales assumed in the PBS.  
Furthermore, the amplitude of the non-Gaussian halo bias is determined
by different quantities in both approaches, and these quantities do not in 
general agree outside a specific high-peak limit of local biasing.

The paper is structured as follows:  we review the case of general,
quadratic primordial non-Gaussianity in \refsec{NG}.  \refsec{PBS}
presents the peak-background split argument, and \refsec{bias} derives
the non-Gaussian halo bias in this approach.  \refsec{squeezed} contains
some results on the squeezed limit of different primordial bispectra.  
Finally, we compare the PBS and local-biasing approaches in \refsec{PBSvslocal},
and conclude in \refsec{concl}.  An Appendix shows that the
local-model bispectrum can arise from a non-local model, a
mathematical curiosity that arose during the development of the
formalism in this paper.

\section{General quadratic non-Gaussianity}
\label{sec:NG}

\subsection{Basic formalism}

Suppose the primordial potential $\Phi$ is a general, non-local
quadratic function of a Gaussian field $\phi$.  Expressed in
real space, the most general expression preserving statistical
homogeneity and isotropy is a two-dimensional convolution,
\be
     \Phi(\vx) = \phi(\vx) + \fNL\int d^3y\int d^3z\;
     W(\vy,\vz) \phi(\vx+\vy)\phi(\vx+\vz),
\label{eq:Phi-real}
\ee
where the kernel $W(\vy, \vz)$ is only a function of $y$, $z$, and 
$\v{\hat{y}}\cdot\v{\hat{z}}$ and is symmetric,
$W(\vy,\vz)=W(\vz,\vy)$.  Here, the non-Gaussianity amplitude is
quantified by the parameter $\fNL$.  Although the spatial
average of $\Phi$ is not zero, but of order $\<\phi^2\>$, this
constant offset is unobservable and can be eliminated by a gauge
transformation.

In Fourier space, Eq.~(\ref{eq:Phi-real}) reads
\ba
     \Phi(\vk) = \phi(\vk) + &\fNL \int
     \frac{d^3k_1}{(2\pi)^3} \label{eq:Phi}\\ & \times
     \Wt(\vk_1, \vk-\vk_1) \phi(\vk_1) \phi(\vk-\vk_1),\nonumber
\ea
where $\Wt$ is the Fourier transform,
\be
     \Wt(\vk_1,\vk_2) = \int d^3y \int d^3z\:
     e^{-i\vk_1\cdot \vy-i\vk_2\cdot\vz} W(\vy,\vz).
\label{eq:Fouriertransform}
\ee
Statistical homogeneity requires that $\Wt$ has no dependence
on the directions of $\vk_1$ and $\vk_2$, only on their relative
directions.  It is thus a function only of the magnitudes $k_1$
and $k_2$ (and symmetric in these arguments) and the dot product
$\vk_1 \cdot \vk_2$.  It can thus alternatively be written as a
function $\Wt(k_1,k_2,k_3)$ of the magnitude $k_3=|\vk_1+\vk_2|$
of the third side of the triangle constructed from $\vk_1$ and
$\vk_2$.

The leading non-Gaussian correction to the potential is
commonly parametrized in terms of the bispectrum
$B(k_1,k_2,k_3)$ of $\Phi$, defined by,
\be
     \<\Phi(\vk_1)\Phi(\vk_2)\Phi(\vk_3)\> =
     \;(2\pi)^3 \d_D(\vk_1+\vk_2+\vk_3) B_\Phi(k_1,k_2,k_3),
\ee
where $\d_D$ is a Dirac delta function.
It is straightforward to calculate the bispectrum of $\Phi$
given Eq.~(\ref{eq:Phi}).  We obtain
\begin{eqnarray}
     B_\Phi(k_1,k_2,k_3) = 2\fNL \bigg\{ &
     \Wt(\vk_1,\vk_2) P_\Phi(k_1) P_\Phi(k_2) \vs
& + 2\:\perm
\bigg\}.
\label{eq:B-Wt}
\end{eqnarray}
Here, $P_\Phi(k)$ denotes the power spectrum of $\Phi$, which to leading order
in $\fNL$ agrees with that of $\phi$.  The two permutations not
written are the two remaining cyclic permutations of
$k_1,k_2,k_3$.   The requirement of statistical isotropy
dictates that $B_\Phi$ is a function only of the triangle shape,
not its orientation, and so the bispectrum can be written as a
function $B_\Phi(k_1,k_2,k_3)$ only of the magnitudes of its
three arguments.  The Fourier-space kernel $\Wt$ is, on the
other hand, only required to be symmetric under exchange of its
two (vectorial) arguments.  Hence, Eq.~(\ref{eq:B-Wt}) does not
uniquely define $\Wt$; there may be several different $\Wt$ that
yield the same bispectrum (note however that these will give different
trispectra at order $\fNL^2$).

The ambiguity in $\Wt$ can be eliminated if we assume, for
example, that it is symmetric under exchange of its three
arguments.  With this additional assumption, the Fourier-space
kernel can be written in terms of the bispectrum as
\be
     \Wt_{\mathrm{sym}}(k_1,k_2,k_3) \equiv \frac{1}{2\fNL}
     \frac{B_\Phi(k_1,k_2,k_3)}{ P_1 P_2 + P_1 P_3 + P_2 P_3},
\label{eq:Wtilde}
\ee
where $P_i\equiv P_\Phi(k_i)$.  Eq.~(\ref{eq:Wtilde}) becomes more
clear by considering concrete examples.  In the local model,
the bispectrum is given by \cite{localmodel,Luo:1993xx,Verde:1999ij}
\be
     B^{\rm loc}_\Phi(k_1,k_2,k_3) = 2\fNL\left[ P_\Phi(k_1)
     P_\Phi(k_2) + 2\;\perm\right].
\label{eq:localmodelbispectrum}
\ee
Applying Eq.~(\ref{eq:Wtilde}) immediately yields
\be
     \Wt^{\rm loc}_{\mathrm{sym}}(k_1,k_2,k_3) = 1,
\ee
and going back to real-space, we obtain 
\be
     W^{\rm loc}(\vy,\vz) = \d_D(\vy) \d_D(\vz).
\ee
This, together with Eq.~(\ref{eq:Phi-real}), yields the
well-known local expression for $\Phi$,
\be
     \Phi(\vx) = \phi(\vx) + \fNL \phi^2(\vx).
\label{eq:localmodelPhi}
\ee
Finally, we note that the prediction of the scale-dependent non-Gaussian
halo bias on large scales does not depend on the particular choice of 
kernel.  As we
will see in \refsec{bias}, this prediction depends on the squeezed limit
of the kernel, where $k_2 \gg k_1$.  Denoting $q = k_1/k_2$, this limit
corresponds to $q\rightarrow 0$, and in the same limit \refeq{B-Wt} can
be written as\footnote{This assumes that the kernel in the last
permutation of \refeq{B-Wt}, involving the two large $k$ values, does not 
grow faster than $q^{-3}$.}
\be
\Wt(\vk_1,\vk_2) \stackrel{\rm squeezed}{=} \frac{B_\Phi(k_1,k_2,k_3)}
{4\fNL P_\Phi(k_1) P_\Phi(k_2)} \left ( 1 + \mathcal{O}(q)\right).
\label{eq:Wt-sq}
\ee
Thus, in the squeezed limit the kernel $\Wt$ and in particular its leading scaling
with $q$ are uniquely specified.  Furthermore, \refeq{B-Wt}
also shows that all kernels agree for equilateral triangles, further restricting
the magnitude of the kernel-choice ambiguity.  In \refsec{squeezed} we will 
return to 
the squeezed limit of $\Wt$ and show how it determines the scale-dependence
of the non-Gaussian halo bias.

\subsection{Some primordial bispectra}

We now summarize other non-local bispectra that have been
discussed in the recent literature and that we will consider
here.  All of these bispectra are scale-free [i.e. they 
scale as $P_\Phi^{2}$ under a uniform rescaling of all $k_i$], but we 
reiterate that this is
not necessary for Eqs.~(\ref{eq:Phi}) and (\ref{eq:Wtilde}) to apply.
These bispectra simplify considerably in the squeezed limit, which
we will discuss in \refsec{squeezed}.

In the \emph{equilateral} model \cite{Equil}, the
bispectrum is given by
\ba
     B_\Phi^{\rm eql} = 6 \fNL \big [&  (-P_1 P_2 - 2\:\perm) - 2
     (P_1 P_2 P_3)^{2/3}\vs & + (P_1^{1/3} P_2^{2/3} P_3 +
     5\:\perm) \big].
\label{eq:Beql}
\ea
The \emph{folded} bispectrum is given by \cite{folded}
\ba
     B_\Phi^{\rm fol} = 6 \fNL \big [& (P_1 P_2 + 2\:\perm) + 3
     (P_1 P_2 P_3)^{2/3}\vs & - (P_1^{1/3} P_2^{2/3} P_3 + 5\:\perm) \big ].
\label{eq:Bfol}
\ea
The \emph{orthogonal} bispectrum is given by \cite{Senatore:2009gt}
\ba
   B_\Phi^{\rm orth} = 6 \fNL \Big\{& -3\left( P_1 P_2
     + 2\, \mathrm{perms}\right) - 8 (P_1 P_2 P_3)^{2/3}  \nonumber \\ 
&   + 3 \left[ (P_1 P_2^{1/3} P_3^{2/3}) + 5\,\mathrm{perms} \right] \Big\}.
\label{eq:Borth}
\ea
The bispectrum due to self-ordering scalar fields (SOSFs) is
\cite{Figueroa:2010zx}
\begin{eqnarray}
     B_\Phi^{\rm SOSF} &=& 437\, \fNL \left(P_1 P_2 P_3
     \right)^{2/3} \left(1-0.485 \frac{k_2}{k_1} \right)
     \nonumber \\
     & & \times \frac{k_3}{k_1} \left[ 1-1.87 \frac{k_3}{k_1} +
     0.945 \left(\frac{k_3}{k_1} \right)^2 \right],
\label{eq:SOSFbispectrum}
\end{eqnarray}
where in this case we restrict $k_1\geq k_2 \geq k_3$.  We adopt
the usual convention of defining $\fNL$ so that $B_\Phi= 6\fNL
(P_\Phi)^2$ for equilateral triangles ($k_1=k_2=k_3$), except
for the folded bispectrum, which is zero for equilateral
triangles.

\subsection{Relating the processed and primordial power spectra}

So far we have been discussing the {\it primordial} potential
and its power spectrum and bispectrum, but what we need for
galaxy clustering is the {\it processed} potential.  The
processed potential, which we denote with a subscript ``0,'' is
related to the primordial potential in Fourier space via
$\Phi_0(\vk) = T(k) \Phi(\vk)$, where $T(k)$ is the transfer
function.  If we similarly define a processed Gaussian random
field $\phi_0(\vk)=T(k)\phi(\vk)$, then the processed potential
may be written,
\ba
     \Phi_0(\vk) = \phi_0(\vk) + &\fNL \int
     \frac{d^3k_1}{(2\pi)^3} \label{eq:Phi0}\\ & \times
     \Wt_0(\vk_1, \vk-\vk_1) \phi_0(\vk_1)
     \phi_0(\vk-\vk_1),\nonumber
\ea
in terms of a processed kernel,
\begin{equation}
     \Wt_0(\vk_1,\vk_2) = \frac{ T(|\vk_1 + \vk_2|) \Wt(\vk_1, \vk_2)}
     {T(k_1) T(k_2)}.
\label{eq:Wt0}
\end{equation}
Note that even if we choose $\Wt(\vk_1,\vk_2)$ to be
a symmetric function $\Wt_{\mathrm{sym}}(k_1,k_2,k_3)$ of the three magnitudes
$k_1$, $k_2$, $k_3$, the processed kernel, $\Wt_0(k_1,k_2,k_3)$
is {\it not} symmetric under exchange of all three arguments.  Still,
it is symmetric under exchange of its two vectorial arguments
if written $\Wt_0(\vk_1,\vk_2)$.  The configuration-space kernel
$\Wt_0(\vy,\vz)$ is then obtained from the Fourier transform of
$\Wt_0(\vk_1,\vk_2)$.  It is this processed kernel
that will appear in the analysis below.

Alternatively, one can define the processed kernel $\Wt_0$
directly from the bispectrum and power spectrum of the processed
potential $\Phi_0$, in analogy with \refeq{Wtilde}:
\begin{equation}
     \Wt_0(\vk_1,\vk_2) =
     \frac{1}{2\fNL}\frac{B_{\Phi_0}(k_1,k_2,|\vk_1 + \vk_2|)}
     {P_{\Phi_0}(k_1) P_{\Phi_0}(k_2) + {\rm cyc.}},
\label{eq:Wt0-alt}
\end{equation}
where 
\ba
     B_{\Phi_0}(k_1,k_2,k_3) =\:& T(k_1) T(k_2) T(k_3) B_\Phi(k_1,k_2,k_3),\\
     P_{\Phi_0}(k) =\:& T^2(k) P_\Phi(k).
\ea
Note, however, that if we choose the {\it processed} kernel to
be symmetric in all three arguments, then the {\it primordial}
kernel will no longer be symmetric.  That is, either the primordial
or the processed kernel may be fully symmetric, but not both.

In our analysis below, we will consider these two forms for
$\widetilde W_0$, one (which we label with the subscript ``1'') in
which the primordial kernel is fully symmetric, and one (which
we label with the subscript ``2'') in which the processed kernel
is symmetric.  To be clear, the two kernels we consider below are
\begin{equation}
     \Wt_{01}(\vk_1,\vk_2) \equiv \frac{T(|\vk_1+\vk_2|)}{T(k_1)
     T(k_2)} \widetilde W_{\mathrm{sym}}(\vk_1,\vk_2),
\label{eq:Wt01}
\end{equation}
and
\begin{equation}
     \Wt_{02}(\vk_1,\vk_2) \equiv
     \frac{1}{2\fNL}\frac{B_{\Phi_0}(k_1,k_2,|\vk_1 + \vk_2|)}
     {P_{\Phi_0}(k_1) P_{\Phi_0}(k_2) + {\rm cyc.}}.
\label{eq:Wt02}
\end{equation}
There are, of course, other kernels one could
choose.  However, with any definition, $\Wt_0 = \Wt$ whenever
all three $\vk$ vectors are sufficiently small so that $T(k)\simeq
1$.  Moreover, they have the same squeezed limits:
\ba
\Wt_0(\vk_1,\vk_2) \stackrel{k_2 \gg k_1}{\longrightarrow} &
\frac{1}{T(k_1)} \Wt(\vk_1,\vk_2) \label{eq:Wt0squeezed}\\
=\:& \frac{1}{T(k_1)} \frac{B_\Phi(k_1,k_2,k_3)}
{2\fNL P_\Phi(k_1) P_\Phi(k_2)}, \nonumber
\ea 
where we have used \refeq{Wt-sq} for the last equality.  This shows
that the processed kernel is uniquely specified in the squeezed limit
as well.  Further, $\Wt_{01}$ and $\Wt_{02}$ also have
the same equilateral limits, which further restricts the
numerical uncertainty associated with this theoretical
ambiguity.  Finally, note that $\Wt_0 \neq 1$ in general for the local
model, even though $\Wt$ is unity.  This reflects the fact that
the relation between the primordial and processed potentials (and
matter perturbations) is non-local.

Before proceeding, we note that one consequence of our approach
is the realization that a non-local inflationary model can give rise to the
local-model primordial bispectrum.  This is spelled out in the
Appendix.

\section{Peak-background split for general quadratic non-Gaussianity}
\label{sec:PBS}

The physical, non-Gaussian, matter-density perturbation $\dng$ is related 
to the processed potential $\Phi_0$ through the Poisson equation,
\be
\frac{\nabla^2}{a^2} \Phi_0 = 4\pi G\:\bar\rho\:\dng.
\label{eq:Poisson}
\ee
Similarly, we can define a fictitious Gaussian density field
$\d$ related to the Gaussian potential $\phi_0$ by the same
relation.   In Fourier space, these relations read,
\bea
     \dng(\vk,z) &=& \frac{2}{3}\frac{k^2}{(1+z)\,H_0^2 \Om}
     g_*(z) \Phi_0(\vk,z_*)\vs
     &\equiv& \M(k,z) \Phi_0(\vk,z_*),\label{eq:d-Phi}\\
     \d(\vk,z) &=& \M(k,z) \phi_0(\vk, z_*),
\eea
where $g_*(z) \propto (1+z) D(z)$ is the potential growth factor normalized to 
unity at last scattering (at which in our convention $\fNL$ is defined).  

In the peak-background split, we first divide any perturbation $A(\vx)$ into a
short- and long-wavelength piece:
\be
A(\vx) = A_s(\vx) + A_l(\vx).
\ee
Here, the ``long'' wavelengths are those comparable to the
distances over which we measure correlations.  The ``short''
wavelengths are on much smaller scales, of order the Lagrangian extent
of halos and less, whose peaks
determine the locations of halos.  
For the Gaussian potential $\phi_0$, the two pieces $\phi_l \equiv \phi_{0,l}$ 
and $\phi_s \equiv \phi_{0,s}$
are uncorrelated.  The non-Gaussianity of Eq.~(\ref{eq:Phi}) however induces a
coupling of long- and short-wavelength perturbations in the potential
as well as the matter density.  Using Eq.~(\ref{eq:Phi-real})
and the Poisson equation, \refeq{Poisson},
we can write the density $\dng$ in terms of the Gaussian long- and
short-wavelength pieces,
\bea
     \dng(\vx) &=& \d_l(\vx)+\d_s(\vx) + 2\fNL \int d^3x\int
     d^3y\; W_0(\vy,\vz) \vs 
      & & \times\bigg \{ \phi_l(\vx+\vy) \left[\d_l(\vx+\vz) +
      \d_s(\vx+\vz)\right]\vs
       & &\quad\   + \phi_s(\vx+\vy) \left[\d_l(\vx+\vz) +
       \d_s(\vx+\vz)\right]\vs 
        & &\quad\   - \frac{1}{2}\vn_x\phi_0(\vx+\vy)\cdot\vn_x\phi_0(\vx+\vz)
        \bigg \}.
\eea
We are interested in the density field at the positions of density peaks.  
In those locations, we expect terms of the form $\phi_0\:\d$ to be much larger
than $(\vn\phi_0)^2$, since $\d$ is at least order unity, and additionally
$\vn\phi_0$ is suppressed if we associate peaks in the density field
with peaks in the potential (where $\vn\phi_0 = 0$). 

Next, we are mainly interested in the non-Gaussian effects on small-scale
modes $\dng_s$, since those result in the leading effect on halo formation.  
Separating the purely long-wavelength parts, we have
\bea
\dng_s(\vx) &=& \d_s(\vx) + 2\fNL \int d^3x\int d^3y\; W_0(\vy,\vz)  \vs
& &\times \bigg \{  \phi_l(\vx+\vy)\d_s(\vx+\vz) \\
& &\quad\   + \phi_s(\vx+\vy) \left[\d_l(\vx+\vz) + \d_s(\vx+\vz)\right] \bigg \}.\nonumber
\eea
We will now neglect the terms in the last line: $\phi_s\d_l$ is very small,
since $\phi_s\sim\phi_l$, and $\d_l \ll \d_s$; and $\phi_s\d_s$, while large, 
adds a purely stochastic contribution on large scales.  We then obtain the
final expression for the effect of primordial non-Gaussianity on small-scale
modes in the context of the peak-background split:
\ba
\dng_s(\vx) =\;& \d_s(\vx) \label{eq:deltas-real}\\
+ 2\fNL &\int d^3x\int d^3y\; W_0(\vy,\vz)
\phi_l(\vx+\vy)\d_s(\vx+\vz).\nonumber
\ea
Equivalently, in Fourier space:
\ba
     \dng_s(\vk_s) =\;& \d_s(\vk_s) \label{eq:deltas}\\
      + 2\fNL &\int \frac{d^3k}{(2\pi)^3} 
      \Wt_0(\vk,\vk_s-\vk) \phi_l(\vk) \d_s(\vk_s-\vk).\nonumber
\ea
Note that by assumption, $k_s \gg k$: the effect on the small-scale modes
is determined by the behavior of the bispectrum in the squeezed limit
[Eq.~(\ref{eq:Wtilde})].  In the following, we denote short-wavelength modes
with $k_s$, while $k$ will stand for long-wavelength modes.  We reiterate
that even for the local model, the coupling between density and potential
is non-local, since $\Wt_0 \neq 1$.

\section{First-order, non-local halo bias}
\label{sec:bias}

We assume that the number density of halos, per unit logarithmic
mass, smoothed over a region of size $\lambda \gtrsim k_s^{-1}$
centered at position $\vx$ is a function,
\be
     n = n\Bigl(M,z;\rhob\left[1+\delta_l(\vx) \right],P\bigl(k_s,\delta_l(\vx)
     \bigr) \Bigr),
\label{eq:massfunction}
\ee
where $M$ is the halo mass, $z$ the redshift,
$\bar\rho\:[1+\delta_l(\vx)]$ is the density of matter in that region,
and $P(k_s,\delta_l(\vx))$ is the small-wavelength (linear) matter power
spectrum in that region.  In a smooth Universe---i.e., one with no
long-wavelength fluctuations, $\delta_l=0$, the halo abundance
is the same everywhere.  But if
$\delta_l \neq0$, then there
will be fluctuations in the halo abundance with long-range
correlations determined by long-range correlations in the matter
density.  Note that if primordial perturbations are Gaussian,
then $P(k_s)$ is the same everywhere.  If, however, primordial
perturbations are non-Gaussian, and if that non-Gaussianity
couples short- and long-wavelength modes of the density field,
then the small-scale power spectrum $P(k_s,\delta_l(\vx))$ may
vary from one point in the Universe to another, as written in
Eq.~(\ref{eq:massfunction}).

Biasing describes the relative clustering of halos and matter.
The Lagrangian bias $b_L$ is defined to be the ratio of the
fractional halo- and matter-density perturbations.  More
precisely, here we will consider the scale-dependent bias
$b_L(k)$ which describes the relative amplitudes of Fourier
modes of the halo abundance and matter density. To proceed,
consider a single Fourier mode $\delta_l(\vx) = \tilde\delta_l(k)
e^{i\vk \cdot \vx}$ of the density field with wavevector $\vk$
and amplitude $\tilde \delta_l(\vk)$.  This Fourier mode induces a
variation $\delta n(\vx) = \delta \tilde n(\vk) e^{i\vk\cdot
\vx}$ with amplitude $\delta \tilde n(\vk) =
(dn/d\tilde\delta(\vk))_{\vx=0,\delta_l=0}\; \tilde\delta_l(\vk)$.  The
scale-dependent Lagrangian bias is thus,
\begin{eqnarray}
     b_L(M,z;k) &\equiv& \frac{ \delta \tilde n(\vk)/\bar n}{\delta
     \tilde\rho/\bar\rho} = \frac{ d \ln \tilde n(\vk)}{d \tilde
     \delta_l(\vk)} \label{eq:sdbias} \\
     &=&  \frac{\partial \ln \tilde n(\vk)}{\partial
     \delta_l(\vk)} + \sum_{\vk_s} \frac{\partial \ln \tilde
     n(\vk)}{\partial P(k_s)} \frac{\partial P(k_s)}{\partial \tilde
     \delta_l(\vk)}. \nonumber
\end{eqnarray}
Here, the wavenumbers $k$ of interest are for long-wavelength
modes, those over which halo clustering is measured, and we have
used the chain rule in the last line.

The first term in the last line of Eq.~(\ref{eq:sdbias}) is the
standard result, that obtained assuming Gaussian initial
conditions.  It evaluates to
\be
     b_L^{\mathrm G} (M,z) = \frac{1}{\bar n}\frac{\partial
     n(M,z;\rhob,P)}{\partial\ln \rhob} - 1.
\label{eq:bPBS}
\ee

The second term in the last line of Eq.~(\ref{eq:sdbias}) arises if
there are non-Gaussian initial conditions that correlate long-
and short-wavelength modes.  In this case, a change in
$\tilde\delta_l(\vk)$ induces changes in $P(k_s)$ for each
$\vk_s$ (hence the sum in Eq.~(\ref{eq:sdbias})) which then
induce changes in $\tilde n(\vk)$.  

To evaluate the partial
derivatives $\partial P(k_s)/\partial \tilde \delta_l(\vk)$, we
multiply Eq.~(\ref{eq:deltas-real}) by $\d_s(\vx')$, letting $\vr = \vx-\vx'$,
and take the expectation value.  The first term that arises is
the usual Gaussian two-point correlation function $\xi_s(r)$.
The non-Gaussian correction---that which correlates short- and
long-wavelength modes---is then
\bea
     \D\xi_s(r) &=& 4\fNL \int d^3y\int d^3z\; W_0(\vy,\vz)
     \phi_l(\vy)\xi_s(\vr+\vz),\quad\quad
\eea
where we have set $\vx=0$ without loss of generality.  We now use
\bea
     \D\xi_s(r) &=& \int\frac{d^3k_s}{(2\pi)^3} e^{i \vk_s
     \vr} \D P(k_s).\label{eq:xi-P}
\eea
Expressing all quantities by their Fourier transforms, and performing the
$\vy,\vz$ integrals yields,
\ba
     \D\xi_s(r) =\;& 4\fNL \int\frac{d^3k_s}{(2\pi)^3} e^{i
     \vk_s \vr} \int\frac{d^3k_1}{(2\pi)^3}\vs 
     & \times  \Wt_0(\vk_1,\vk_s-\vk_1) \phi_l(\vk_1)
     P(k_s).
\ea
Hence, via Eq.~(\ref{eq:xi-P}), the effect of any quadratic primordial
non-Gaussianity on the statistics of the short-wavelength modes
is given by
\be
     \D P(k_s) = 4\fNL \int\frac{d^3k_1}{(2\pi)^3}
     \Wt_0(\vk_1, \vk_s-\vk_1) \phi_l(\vk_1) P(k_s).
\label{eq:DPs}
\ee
From this, we obtain the required partial derivatives,
\ba
\frac{\partial P(k_s)}{\partial\tilde \d(\vk)} =\:& \M^{-1}(k)
     \frac{\partial \D P(k_s)}{\partial \phi_l(\vk)},\label{eq:dPs}\\
\frac{\partial \D P(k_s)}{\partial \phi_l(\vk)} =\:& 4\fNL P(k_s)
     \Wt(\vk,\vk_s-\vk),\nonumber
\ea
recalling that $\phi_l$ refers to the processed potential.  
Replacing the sum $\sum_{\vk_s}$ by an integral $\int d^3k_s/(2\pi)^3$,
we then obtain for the second term in the last line of
Eq.~(\ref{eq:sdbias}) the non-Gaussian correction to the Lagrangian
bias,
\begin{eqnarray}
     \D b_L(M,z;k) &=& 2\fNL\M^{-1}(k) \nonumber \\
     &  \times & \int \frac{d^3k_s}{(2\pi)^3} 2\frac{\partial\ln
     n}{\partial P(k_s)} \Wt_0(\vk, \vk_s-\vk) P(k_s). \nonumber\\
\label{eq:Db-1}
\end{eqnarray}

Thus, in general, non-Gaussianity changes the shape of 
$P(k_s)$ as well as the amplitude.  However, Eq.~(\ref{eq:Db-1}) simplifies
for a very broad class of mass-function models, which only depend on
the power spectrum through an overall normalization.  We will turn to this
case next.

\subsection{Universal mass functions}
\label{sec:univ}

Most of the commonly used parametrizations of the halo abundance
$n(M,  z;\rhob,P)$ are so-called \emph{universal mass
functions} \cite{universal,Verde:2000vq}.  They can be written as
\be
n = \frac{\rhob}{M} \nu f(\nu) \frac{d\ln \nu}{d\ln M},
\label{eq:n_univ}
\ee
and thus only depend on cosmology through the background density
of matter $\rhob$ and the quantity,
\be
\nu = \frac{\d_c^2}{\s^2_R(z)},
\label{eq:nu}
\ee
where $\d_c \approx 1.686$ is the linearly extrapolated collapse threshold.  
Here, $\s_R^2$ is the variance of the density field on a scale
$R$ related to $M$ by $M=4\pi/3\rhob R^3$,
\be
     \s_R^2 = \int\frac{d^3k}{(2\pi)^3} F_R^2(k) P(k),
\label{eq:sR}
\ee
where $F_R$ is the Fourier transform of a real-space tophat filter.  The
parametrization, \refeq{n_univ}, is sufficient to obtain the Gaussian and 
non-Gaussian halo bias.  

First, for the Gaussian halo bias we use
Eq.~(\ref{eq:bPBS}).  
The effect of a long-wavelength perturbation $\d_l$ on the collapse threhold
is $\d_c \rightarrow \d_c-\d_l$, as derived in
Refs.~\cite{Mo:1995cs,Sheth:1999mn}.  This gives (see also
Ref.~\cite{Slosar:2008hx}),
\be
b_L = \frac{\partial\ln n(\rhob,\nu)}{\partial \d_l} = -\frac{2\nu}{\d_c} 
\frac{d\ln ( \nu f(\nu))}{d\nu}.
\ee

On the other hand, it is clear from Eqs.~(\ref{eq:Db-1}) and
(\ref{eq:n_univ}) that the non-Gaussian halo bias will only enter through
\be
\frac{\partial\ln n(\rhob,\nu)}{\partial\ln \s_R} = -2\nu
\frac{d\ln (\nu f(\nu))}{d\nu} = b_L \d_c.
\label{eq:dnds}
\ee
Next, we again use the chain rule, and, from \refeq{sR},
\be
\frac{\partial \s_R^2}{\partial P(k_s)} = F_R^2(k_s).
\ee
We then have
\ba
\D b_L(k) =\;& 2\fNL \M^{-1}(k) \frac{\partial\ln n}{\partial \ln \s_R}
\frac{\s_W^2(k)}{\s_R^2},\label{eq:Db-1b}\\
\s_W^2(k) \equiv\;& \int \frac{d^3k_s}{(2\pi)^3} P(k_s) F_R^2(k_s)
 \Wt_0(\vk, \vk_s-\vk).\nonumber
\ea
The prefactor $\M^{-1}$ scales as $k^{-2}$.  However, in general $\s_W^2$,
which involves $\Wt_0(\vk_s,\vk)$, will depend on $k$ as well.  

Making use of the bias relations derived above, we have
\be
     \D b_L(k) = 2\fNL \M^{-1}(k) b_L \d_c
     \frac{\s_W^2(k)}{\s_R^2},
\label{eq:Db-2}
\ee
where, to clarify, the matter power spectrum $P(k)$ is related
to the processed-potential power spectrum $P_{\Phi_0}(k)$
through $P(k)= [\M(k)]^2 P_{\Phi_0}(k)$.

Let us first look at the local case.  We use the squeezed limit
[\refeq{Wt0squeezed}] and $\Wt^{\rm loc}=1$.  In this case,
the change in the small-scale power spectrum is independent of $k_s$, so
that only the overall amplitude is affected.  We obtain:
\be
\D b^{\rm local}_L(k) = 2\fNL \frac{1}{\M(k) T(k)} b_L \d_c,
\label{eq:Dbloc}
\ee
exactly matching the results of Refs.~\cite{Slosar:2008hx}.  
We will quantify the regime of validity of the squeezed limit in
\refsec{PBSvslocal}.

For a given bispectrum, Eq.~(\ref{eq:Db-2}) can be evaluated exactly
in a straightforward way.  However, it is also interesting to look at
the scaling of the bispectrum in the squeezed limit, which allows for a 
simpler estimate of $\D b$.

\begin{figure}[t!]
\centering
\includegraphics[width=0.48\textwidth]{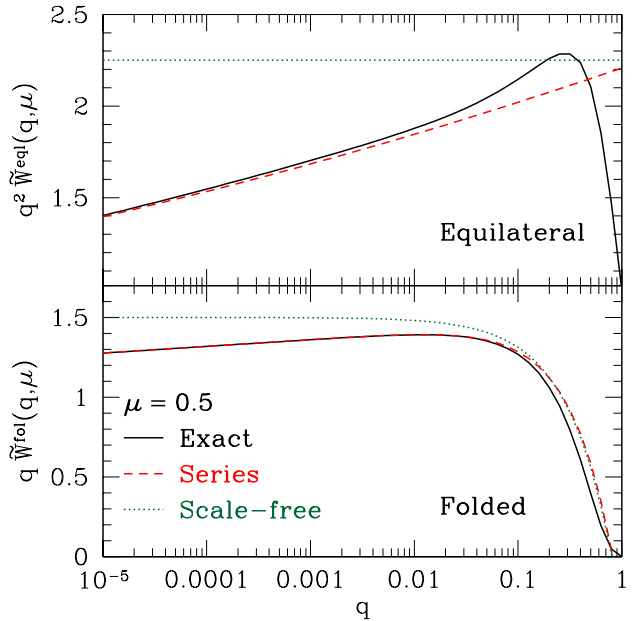}
\caption{Squeezed limit of $\Wt(q,\mu)$ as function of $q$ for 
equilateral (top) and folded (bottom) bispectra.  Here, we have taken out
the leading $q$ scaling and adopted
a fixed value of $\mu=0.5$.  Shown is
the exact result from Eq.~(\protect\ref{eq:Wtilde}), and the
series expansion of the squeezed limit to
$\mathcal{O}(q^2,\eps^2)$ from \refsec{squeezed}.  We also show the
scale-free limit of this series expansion, which in general is not
a good approximation.  
\label{fig:WtSq}}
\end{figure}

\section{Squeezed limit and Scale-Dependence of Bias}
\label{sec:squeezed}

As we have seen, the effect of primordial non-Gaussianity on the halo
bias depends on $\Wt_0(\vk,\vk_s-\vk)$, where $k_s \gg k$.  This
is known as the squeezed limit, since it corresponds to
triangles where two sides are much longer than the third.  In
this Section, we derive the primordial kernel $\Wt(\vk,\vk_s-\vk)$ in the
squeezed limit to show analytically how the non-Gaussian bias
will behave for the various bispectra we consider.  To leading
order, the processed kernel is then given by \refeq{Wt0squeezed}.  
For our numerical results in the next section, we will use the full 
kernels, \refeqs{Wt01}{Wt02}.  For definiteness, we assume a flat
$\Lambda$CDM cosmology throughout, with $n_s=0.958$, $\Omega_m=0.28$,
$h=0.72$, and $\sigma_8=0.8$.  

Let us define $q = k/k_s$ and $\mu =
\v{\hat{k}}\cdot\v{\hat{k}}_s$.   We then write the kernel as a
power series in $q$.  For example, in the simplest case, the
local model, the squeezed limit is just $\Wt = 1$.  For the
other bispectra, we use the $k$ scaling of $P_\Phi$,
\be
     P_\Phi(k) \propto k^{n_s-4} \equiv k^{-3+\eps},\  \eps = n_s-1,
\ee
where $\eps\approx -0.04$ is the deviation from scale-free
initial perturbations.
The squeezed limit of the non-local bispectrum shapes introduced
in \refsec{NG} can be easily derived in the limit of scale-free initial
conditions.  However, we will see that it is important to take into account
a non-zero $n_s-1$.  Since the prefactor in Eq.~(\ref{eq:Db-1}) goes as 
$k^{-2} \propto q^{-2}$, we will expand $\Wt$ up to terms $\propto q^2$.  
We will also expand to second order in $\eps$.

For the equilateral model, this gives,
\ba
\Wt^{\rm eql}(q,\mu) &\stackrel{\rm squeezed}{=}
3 q^2\: \bigg ( 1 - \mu^2 - \frac{2\eps}{3} (\ln q - 2 \mu^2) \vs
&\qquad\qquad\qquad + \frac{\eps^2}{9} (2 \ln^2 q - \mu^2 ) \bigg )\label{eq:WtSq-eql}\\
&\stackrel{\rm scale-free}{=} 3 q^2\:( 1-\mu^2).
\ea
Thus, in the scale-free case, $\Wt^{\rm eql}$ scales exactly as
$k^2/k_s^2$ in the squeezed limit, canceling
the $k^{-2}$ prefactor in Eq.~(\ref{eq:Db-1}) and leading to a scale-independent
correction to the bias.  This is not quite true in reality, since
Eq.~(\ref{eq:WtSq-eql}) contains logarithmic corrections, which can be
quite important for $q \ll 1$ (\reffig{WtSq}).  Note also that $\Wt=0$
for $\mu=1$ in the scale-free case, but is only suppressed by 
$ 2\eps(\ln q - 2)/3 \approx -0.18$ for $q=0.01$.  

For the folded model, we have,
\ba
& \Wt^{\rm fol}(q,\mu) \stackrel{\rm squeezed}{=}
\frac{3}{2} q \Bigg( 1 + \left(4 \mu^2 + \mu -4\right) \frac{q}{2} \vs
&     -\frac{\eps}{3}\big (\ln q + \frac{q}{2}\:[8\mu^2 + \mu + (\mu-8)\ln q] \big)
\label{eq:WtSq-fol}\\
&     + \frac{\eps^2}{18} \bigg (\ln^2 q + \frac{q}{2}\:\big [8 \mu^2 +2 \mu \ln q  + (\mu -16) \ln^2 q \big]\bigg) \Bigg ) \vs
&\stackrel{\rm scale-free}{=} \frac{3}{2}q +\frac{3}{4} q^2\:(4\mu^2 + \mu - 4).
\ea
In the scale-free case, the folded model shows a scaling proportional to $q$, with
corrections going as $q^2$.  \reffig{WtSq} shows that the scale-free case
matches the true squeezed limit somewhat more accurately than in the
equilateral case.
\begin{figure}[t!]
\centering
\includegraphics[width=0.48\textwidth]{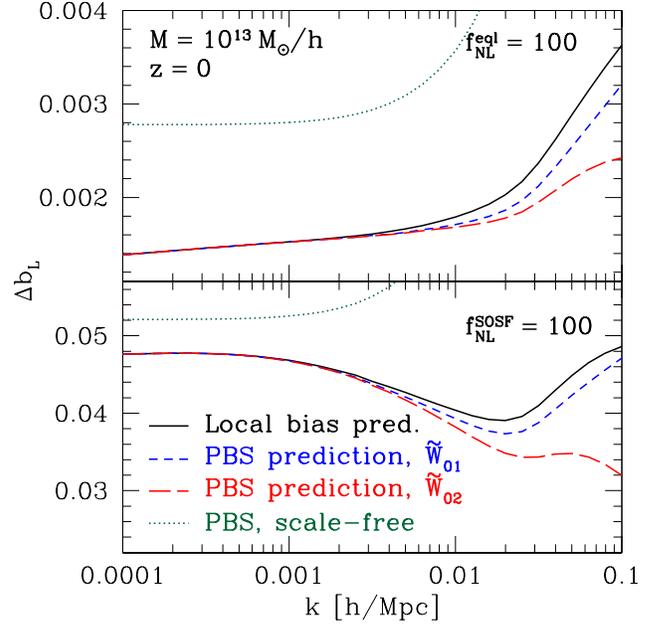}
\caption{Non-Gaussian correction to the halo bias $\D b_L$ calculated
in the peak-background-split (PBS) and local-bias approaches as a function
of $k$, for the equilateral (top) and SOSF (bottom) bispectra.  We have
assumed $\fNL=100$ in both cases and $M=10^{13}\Msunh$ halos at $z=0$.  
We also show the scale-free approximations
Eqs.~(\ref{eq:Dbsf-eql})--(\ref{eq:Dbsf-fol}).  
\label{fig:Deltab}}
\end{figure}

The squeezed limit for the orthogonal model is
\ba
& \Wt^{\rm orth}(q,\mu) \stackrel{\rm squeezed}{=}
-3 q \Bigg( 1 + \frac{q}{2}\left(6\mu^2 + \mu - 6\right) \vs
&     +\frac{\eps}{3}\big (-\ln q + \frac{q}{2}\:[-12\mu^2 - \mu + (12-\mu)\ln q] \big)
\label{eq:WtSq-orth}\\
&     + \frac{\eps^2}{18} \bigg (\ln^2 q + \frac{q}{2}\:\big [12 \mu^2 +2 \mu \ln q  + 
(\mu - 24) \ln^2 q \big]\bigg) \Bigg ) \vs
&\stackrel{\rm scale-free}{=} -3 q + 9 q^2 - \frac{3}{2}q^2 (\mu + 6\mu^2).
\ea
This bispectrum will therefore give rise to
a scale-dependent bias similar to that for the folded
model, but roughly twice as large in magnitude and opposite in sign.

For the SOSF model, the kernel reduces in the squeezed limit to
\ba
     \Wt^{\rm SOSF}(q,\mu) \stackrel{\rm squeezed}{=}
     56.26\:q^2 
     \left ( 1 - \frac{\eps}{3} \ln q \left [ 1 - \frac{\eps}{6}\ln q
     \right]\right). \label{eq:WtSq-sosf}
\ea
The $q^2$ scaling
implies that this model will behave similarly to the equilateral model, 
although the prefactor of 56 indicates that the amplitude of the effect
will be much larger for a given $\fNL$.  

\begin{figure}[t!]
\centering
\includegraphics[width=0.48\textwidth]{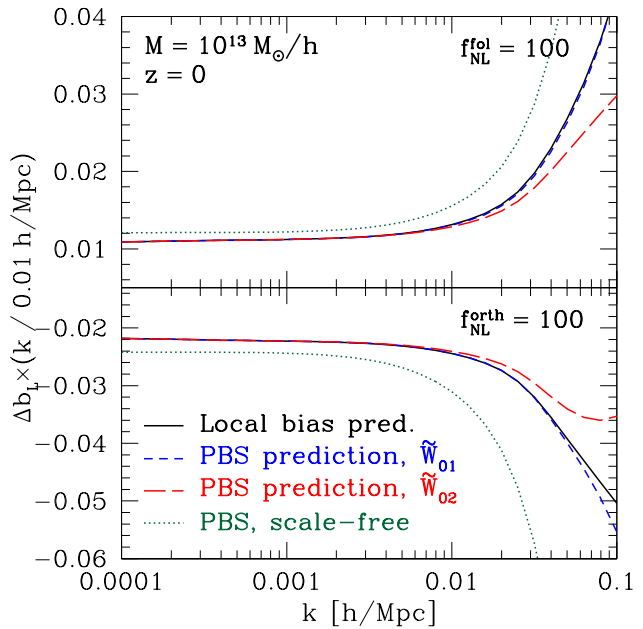}
\caption{Same as \reffig{Deltab}, but for the folded (top) and 
orthogonal (bottom) models.  We have taken out a scaling factor of 
$(k/0.01\iMpch)^{-1}$.  
\label{fig:Deltab-fol}}
\end{figure}

As a rough approximation to the exact result for the non-Gaussian $\D b$,
we can take the scale-free squeezed limit derived above and in addition 
ignore any 
$\mu$-dependence.  From \refeq{Wt0squeezed} and
Eq.~(\ref{eq:Db-2}), we then obtain
\ba
\D b^{\rm eql}_L(k) \approx\;& \frac{6\fNL}{\M(k) T(k)} b_L
\d_c\; k^2 \frac{\s_{R,-2}^2}{\s_R^2},\label{eq:Dbsf-eql}\\
\D b^{\rm SOSF}_L(k) \approx\;& \frac{112\fNL}{\M(k) T(k)}  b_L
\d_c\; k^2 \frac{\s_{R,-2}^2}{\s_R^2},\label{eq:Dbsf-sosf}\\
\D b^{\rm fol}_L(k) \approx\;& \frac{3\fNL}{\M(k) T(k)}  b_L
\d_c\; k \frac{\s_{R,-1}^2}{\s_R^2},\label{eq:Dbsf-fol}\\ 
\D b^{\rm orth}_L(k) \approx\;& \frac{-6\fNL}{\M(k) T(k)}  b_L
\d_c\; k \frac{\s_{R,-1}^2}{\s_R^2},\label{eq:Dbsf-orth}
\ea
where we have defined
\be
\s_{R,n}^2 = \int \frac{d^3k}{(2\pi)^3} k^n\: P(k)\:F_R^2(k).
\ee
Since $\M^{-1}\propto k^{-2}$, we see immediately that $\D b^{\rm eql}$ 
and $\D b^{\rm SOSF}$ are expected to be roughly scale-invariant, while 
$\D b^{\rm fol}$ and $\D b^{\rm orth}$ are expected
to scale as $k^{-1}$ on large scales.  
\reffig{Deltab} shows the exact PBS prediction (in the case of a universal
mass function) for
the non-Gaussian correction $\D b_L$ to the halo bias as a function of $k$, for
the equilateral and SOSF models.  We also show the approximate relations
assuming scale-free initial conditions derived in \refsec{squeezed}.  
We see that this approximation is only good to a factor of a few for these
models.  More importantly, it does not predict the remaining
$k$ dependence of the non-Gaussian correction, which is crucial to disentangle
this effect from the ordinary Gaussian galaxy bias.  The corresponding results
for the folded and orthogonal models are shown in \reffig{Deltab-fol}.  
Here, the scale-free assumption gives a prediction accurate to 
$\sim 20$\% on large scales.  

\section{Peak-background split vs. local biasing}
\label{sec:PBSvslocal}

The effects on the halo power spectrum from non-local non-Gaussianity have
also been derived in a local-Lagrangian-biasing (LLB) scheme
\cite{Verde:2009hy,Matarrese:1986et}.  In this approach,
halos are identified with high-density regions in the initial
\emph{non-Gaussian} density field.  
Specifically, the density $\d_R$ smoothed on a scale $R$ (again related to the
halo mass by $M = 4\pi/3 \rhob R^3$) is required to be greater than the
linear collapse threshold $\d_c$.  This is equivalent to a local-biasing
scheme relating the halo density field $\d_h$ to the smoothed 
(non-Gaussian) matter density field:
\be
\d_h(\vx) = b_L \hat\d_R(\vx) + \frac{b_{L,2}}{2} \hat\d_R^2(\vx) + \dots,
\label{eq:local}
\ee
where $b_{L,2}$ is the second-order Lagrangian bias.  In the high-peak
limit which we will assume here, $b_L = \d_c/\s_R^2$,
and $b_{L,2} = b_L^2$ (see discussion below).  

Since primordial non-Gaussianity induces a non-zero three-point function
in the density field, the large-scale halo correlation function is
modified to
\be
\xi_h(r) = b_L^2 \xi_R(r) + b_L\:b_{L,2}\:\xi_R^{(3)}(\vx_1,\vx_1,\vx_2),
\ee
where $\xi_R$ and $\xi_R^{(3)}$ denote the two- and three-point functions
of the smoothed density field, respectively.  Note that, again, the three-point
correlation function is evaluated in the squeezed limit.  In Fourier space,
this reads \cite{Schmidt:2010pf}
\ba
P_h(k) =\;& b_L^2 P_R(k) + b_L\:b_{L,2}\:\p^{\d\d\d}(k),\\
\p^{\d\d\d}(k) =\;& \int \frac{{\rm d}^3\vk_1}{(2\pi)^3} 
B_R(k, k_1,|\vk_1-\vk|).\label{eq:pddd}
\ea
Here, $P_R(k) \equiv F_R^2(k) P(k)$, and correspondingly for the bispectrum
$B_R$.  Note that in this approach the correction to the
halo power spectrum comes about from the \emph{quadratic} bias parameter.  
With this, we can write the effective correction to the 
\emph{linear} halo bias as
\be
\D b_L(k) = \frac{1}{2} b_{L,2} \frac{\p^{\d\d\d}(k)}{P_R(k)} =
 b_L \d_c \frac{1}{2\s_R^2}\frac{\p^{\d\d\d}(k)}{P_R(k)},
\label{eq:Db-LB}
\ee
where for the second equality we have assumed the high-peak limit.  We
show the predictions of the local-bias model in \reffig{Deltab} for
the equilateral and SOSF models.  
We see that on the largest scales, the peak-background split prediction
agrees precisely with that of local biasing.  On the other hand, on
intermediate scales $k \gtrsim 0.01\iMpch$, the two approaches begin
to diverge, with the peak-background split generally predicting lower
halo bias corrections.  We will now show how this comes about.

\begin{figure}[t!]
\centering
\includegraphics[width=0.48\textwidth]{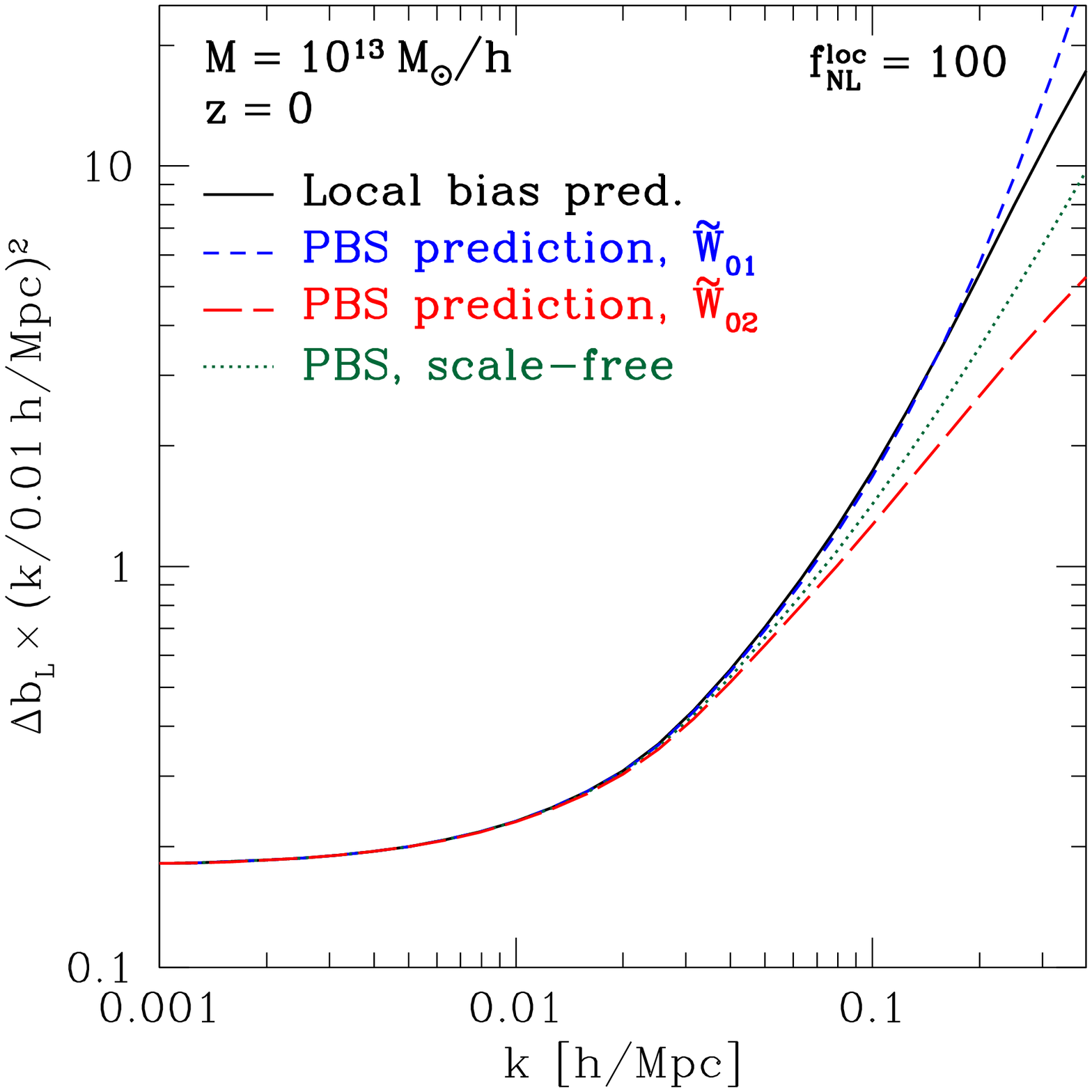}
\caption{Same as \reffig{Deltab}, but now for the local model.  We have
taken out a scaling of $(k/0.01\iMpch)^{-2}$ here.
\label{fig:Deltab-loc}}
\end{figure}

We can write the bias correction in both the PBS and LLB schemes as
(see also \cite{DesjacquesSeljak})
\be
     \D b_L(k) = 2\fNL b_L \d_c \s_R^{-2} \F(k),
\ee
with
\ba
     \F^{\rm PBS}(k) =\;& \frac{1}{\M(k) F_R(k)} \int \frac{d^3k_1}{(2\pi)^3} P_R(k_1)
     \Wt_0(\vk, \vk_1-\vk),\label{eq:FPBS}\\
     \F^{\rm LLB}(k) =\;& \frac{1}{4 \fNL P_R(k)} \int \frac{{\rm
     d}^3\vk_1}{(2\pi)^3}
     B_R(k, k_1,|\vk_1-\vk|).\label{eq:FLB}
\ea
The additional $F_R^{-1}$ prefactor in $\F^{\rm PBS}$ is necessary to match
the local-biasing prediction, since the latter refers the halo bias to
the \emph{smoothed} density field [\refeq{local}]: doing the same in the
PBS approach, via \refeq{sdbias}, \refeq{dPs} and using 
$\tilde\d_R = F_R \tilde\d$, results in this prefactor.

We now insert the definition of $\Wt_0$ [Eq.~(\ref{eq:Wt01}) or
\refeq{Wt02}] into Eq.~(\ref{eq:FPBS}), and use the
relation between the smoothed density field and potential,
\ba
P_R(k) =\;& \M_R^2(k) P_{\Phi_0}(k),\vs
B_R(k_1,k_2,k_3) =\;& \M_R(k_1)\M_R(k_2)\M_R(k_3)
B_{\Phi_0}(k_1,k_2,k_3),\nonumber
\ea
where we have defined $\M_R(k)\equiv \M(k) F_R(k)$.  It is then
straightforward to show that
\ba
\F^{\rm PBS}(k) =\;& \frac{1}{4 \fNL P_R(k)} \int \frac{{\rm d}^3\vk_1}{(2\pi)^3} 
B_R(k, k_1,|\vk_1-\vk|)\vs
& \times A(\vk,\vk_1)\:C(\vk,\vk_1),
\ea
where
\ba
A(\vk,\vk_1) =\;& \frac{\M_R(k_1)}{\M_R(|\vk_1-\vk|)}
 =
 \frac{k_1^2}{|\vk_1-\vk|^2}\frac{F_R(k_1)}{F_R(|\vk_1-\vk|)},
\ea
and
\ba
     C_1(\vk,\vk_1) =\;& 2 \frac{ T^2(k_1)}{T^2(|\vk_1-\vk|)} \vs
     & \times \left(1 + \frac{P_{\Phi}(|\vk_1-\vk|)}{P_{\Phi}(k_1)}
    + \frac{P_{\Phi}(|\vk_1-\vk|)}{P_{\Phi}(k)} \right )^{-1},
\ea
(note that $P_{\Phi}$ is the power spectrum for the
\emph{primordial} potential) if we use $\Wt_{01}$ for the
processed kernel, and
\ba
C_2(\vk,\vk_1) =\;& 2\left(1 + \frac{P_{\Phi_0}(|\vk_1-\vk|)}{P_{\Phi_0}(k_1)}
+ \frac{P_{\Phi_0}(|\vk_1-\vk|)}{P_{\Phi_0}(k)} \right )^{-1},
\ea
(and here $P_{\Phi_0}$ is the power spectrum for the
\emph{processed} potential) if we use $\Wt_{02}$ for the
processed kernel.

In the true squeezed limit, $q = k/k_1$ becomes negligible, so that 
$A \rightarrow 1$ and $C_1\rightarrow 1$ and $C_2\rightarrow 1$
(the second term in the brackets for both expressions
becomes 1, while the last term vanishes in this limit).  Clearly,
$\F^{\rm PBS} = \F^{\rm loc}$ in this limit, which explains why both
approaches agree on large scales.  Going back to Eq.~(\ref{eq:FPBS}), we see
that the integrand peaks where $P_R(k_1)$ peaks, roughly around 
$k_1 \sim 0.02\iMpch$.  Furthermore, as $k_1$ becomes comparable to $k$, 
$|\vk_1-\vk|$ can assume values much smaller than $k_1$.  Thus, we expect that 
deviations of $\F^{\rm PBS}$ from $\F^{\rm LLB}$ appear at least around
$k \sim 0.01\iMpch$, which indeed is visible in \reffig{Deltab}.  Similar
but somewhat smaller deviations occur in the folded and orthogonal cases
(\reffig{Deltab-fol}), and even in the local model of non-Gaussianity 
(see \reffig{Deltab-loc}).  Hence, the agreement between the PBS and LLB 
approaches is restricted to the largest scales in the well-studied local
model as well.  However, since the halo-bias
correction strongly declines towards smaller scales, the
deviations between the two approaches have much less impact in
the local model.  We have confirmed numerically that by introducing the
correction factors $A^{-1}$ and $C^{-1}$ into the PBS expression
\refeq{Db-1b}, we recover the LLB prediction exactly.

On a more physical level, the differences in scale-dependence between the 
PBS and LLB
predictions come from the assumption of a separation of scales between the
large-scale modes modulating the clustering of halos, and the small-scale
modes that govern their formation.  In the course of this assumption, 
several terms in the relation between short-wavelength and long-wavelength
perturbations have been neglected (see \refsec{PBS}).  This assumption breaks down
on smaller scales, however;  modes with $k \sim 0.01\iMpch$, for example,
contribute significantly to the density variance $\s_R^2$, and if we want
to calculate the clustering of halos at that $k$, the separation of scales
does not hold anymore.  
On the other hand, no such assumption of separation of scales is made in
the LLB approach.  Here, the assumption is that halo formation
is purely a function of the local physical density.  
Of course, both approaches need to be benchmarked with
simulations \cite{Wagner:2010me} in the
end, which will have the final say on the scaling of the halo bias
in models with non-local non-Gaussianity.

Even on large scales, there is a difference between the PBS and LLB predictions: 
in the PBS approach, $\Delta b \propto \s_R^{-2}\partial\ln n/\partial\ln\s_R$,
which for any universal mass function equals $b_L\d_c/\s_R^2$.  On the other 
hand, LLB predicts that $\Delta b \propto b_{L,2}$.  While this agrees
with the PBS prediction in the high-peak limit of the Press-Schechter 
theory, where $b_{L,2}= b_L^2 = b_L \d_c/\s_R^2$, this is not the case
in general.  For example, the quadratic bias derived from the Sheth-Tormen
mass function \cite{Sheth:1999mn} in the high-peak limit is 
$b_{L,2} = q\:b_L\d_c/\s_R^2$, thus deviating from 
$\s_R^{-2}\partial\ln n/\partial\ln\s_R$
by a factor of $q \approx 0.75$.  This number will be different in other
mass function prescriptions.  Also, the quadratic 
bias parameter has to change sign at a finite $\nu$, while $b_L\d_c/\s_R^2$
is always positive.  Hence, outside of the high-peak limit the differences
between LLB and PBS predictions can become larger.  It is worth pointing out 
that since there is no prescription for small-scale physics contained in 
the local biasing approach, there is no analogous expression or prediction 
for the quantity $\partial\ln n/\partial\s_R$.  
In the end, a comparison with simulations will have to determine whether the
non-Gaussian bias correction follows a scaling with 
$\partial\ln n/\partial\s_R$, as predicted by the PBS, or with the
quadratic bias $b_{L,2}$, as predicted by local biasing --- both of these
predictions are falsifiable.

\section{Conclusions}
\label{sec:concl}

The principal goal of the theory of large-scale structure is to map
the initial linear density field into the late-time non-linear matter density. 
One of the most fascinating recent applications of this theory is the
effect of primordial non-Gaussianity  on the clustering of dark matter halos.  
Here, we derived predictions for general quadratic non-Gaussianity in the 
peak-background-split (PBS) approach, focusing on the case of
universal halo mass functions.  This complements the derivation
of Ref.~\cite{Verde:2009hy} in the local-bias model.  We also show how
the non-Gaussian effect on halo bias can be calculated beyond universal
mass functions, by taking into account all dependences of the halo abundance
on the statistics of small-scale fluctuations [\refeq{Db-1}].  

While we have only considered general \emph{quadratic} non-Gaussianity
here, it is possible to generalize these results to the case of \emph{cubic}
non-Gaussianity, described by the trispectrum rather than the bispectrum
of primordial fluctuations.  For example, for the local cubic model, we
expect the non-Gaussian halo bias to scale as $\sim k^{-2}$ like in the local
quadratic model \cite{DesjacquesSeljak10b}.  A detailed investigation of cubic non-Gaussianity
is beyond the scope of this paper however.

We show that both PBS and local bias approaches agree in their predictions 
on large scales,
$k\lesssim 0.01\iMpch$, when the high-peak limit is assumed in the local
bias approach, but predictions diverge on smaller scales.  The disagreement
comes about from the breakdown of the separation-of-scales assumption
inherent in the PBS approach.  Even on large scales, however, both approaches
do not in general agree: in the PBS approach (with a universal mass function),
the non-Gaussian effect on the halo bias scales with 
$\partial\ln n/\partial\s_R = b_L\d_c/\s_R$, which quantifies the dependence of
the halo number on the variance of the small-scale fluctuations.  
By the nature of the approach, there is no analogous quantity in the local bias
framework.  Here, the halo bias correction scales with the quadratic
bias parameter $b_{L,2}$ which in general is not directly related to
$\partial\ln n/\partial\s_R$.  This disagreement will also have consequences
for predictions of higher order statistics (e.g., the bispectrum) of halos, 
where both $b_{L,2}$ and the non-Gaussian bias contribution appear.  

N-body simulations will have the final say on which of the
two approaches has the right scaling with halo mass (or equivalently, 
significance $\nu$), and with scale;  it is possible that PBS fares better
in the former respect, while LLB is closer to reality in the latter.

\vspace*{0.5cm}
\begin{acknowledgments}
We would like to thank Neal~Dalal, Olivier~Dor\'e, Lam~Hui, 
Donghui~Jeong, Rom\'an~Scoccimarro, and Sarah~Shandera for 
enlightening discussions.  
This work was supported by DoE DE-FG03-92-ER40701, NASA
NNX10AD04G, and the Gordon and Betty Moore Foundation.
\end{acknowledgments}

\vspace*{0.5cm}
\appendix

\section{The local-model bispectrum from a nonlocal model}
\label{sec:appendix}

In \refsec{NG}, we showed that the bispectrum
$B_{\Phi}(k_1,k_2,k_3)$ does not uniquely determine the kernel
$\Wt(\vk_1,\vk_2)$.  Several different kernels may give rise to
the same bispectrum.  We moreover constructed explicitly two
different \emph{processed} kernels $\Wt_0(\vk_1,\vk_2)$ for a
given bispectrum.  Let us now use Eq.~(\ref{eq:Wt02}) for the
processed kernel for the (processed) bispectrum for the local
model.  If we then invert Eq.~(\ref{eq:Wt0}) to obtain the
primordial kernel that corresponds to this processed kernel, we
find,
\begin{equation}
     \Wt(\vk_1,\vk_2) = \frac{ \left[T(k_1)T(k_2)\right]^2 (P_1 P_2 +
     2\, \mathrm{perms})}{\left[T(k_1)T(k_2)\right]^2 P_1 P_2 +
     2\,\mathrm{perms}}.
\label{eq:alternativelmW}
\end{equation}
Upon substituting into Eq.~(\ref{eq:B-Wt}), we recover the
local-model bispectrum, Eq.~(\ref{eq:localmodelbispectrum}).

Moreover, the function $T(k)$ that appears in
Eq.~(\ref{eq:alternativelmW}) need not be the transfer function
for the local-model bispectrum to be recovered.  Indeed, any
kernel,
\begin{equation}
     \Wt(\vk_1,\vk_2) = \frac{ \left[f(k_1)f(k_2)\right]^2 (P_1 P_2 +
     2\, \mathrm{perms})}{\left[f(k_1) f(k_2)\right]^2 P_1 P_2 +
     2\,\mathrm{perms}},
\end{equation}
for any function $f(k)$ will yield the local-model bispectrum.
And for any function $f(k)$ that is not simply a constant, the
configuration-space kernel $W(\vy,\vz)$ obtained through the
inverse transform of Eq.~(\ref{eq:Fouriertransform}) will be
nonlocal.

In fact, the ambiguity
in the kernel $\Wt$ is much more general: for any positive function
of three arguments, $g(k_1, k_2, k_3)$, and any bispectrum $B_\Phi$, 
we can construct a kernel satisfying \refeq{B-Wt}:
\begin{align}
\Wt(k_1,k_2,k_3) =\:& \frac{g(k_1,k_2,k_3)}{g(k_1,k_2,k_3) P_1 P_2 + 2\,\mathrm{perms}}\nonumber\\
&\:\times \frac{B_\Phi(k_1,k_2,k_3)}{2\fNL}.
\end{align}

We have therefore shown, by explicit construction, that there
are models in which the potential $\Phi$ is a nonlocal quadratic
function of a Gaussian field $\phi$ that have the same
bispectrum as that for the local model. We therefore 
conclude that measurement of the local-model bispectrum does not
necessarily imply that the potential has the local-model form
given in Eq.~(\ref{eq:localmodelPhi}).  Note, though, that these
different models may still differ in their predicted
$\mathcal{O}(\fNL^2)$ trispectra.  For now, we treat this
simply as a mathematical curiosity, and we leave the
investigation of the implications of this result for model
building for future work.


\end{document}